\documentstyle[twocolumn,float,epsfig,prl,aps]{revtex}
\long\def\wideabs#1{\twocolumn[\hsize\textwidth\columnwidth\hsize%
\csname @twocolumnfalse\endcsname #1 \vskip1pc]}

\newcommand{\mpme}{$M_{\pi \mu e}$}

\newcommand{\Kpmel}{$K^{+}\rightarrow\pi^{+}\mu^{+}e^{-}$}
\newcommand{\Kpme}{$K_{\pi \mu e}$}
\newcommand{\kpeel}{$K^{+}\rightarrow\pi^{+}e^{+}e^{-}$}

\newcommand{\kppp}{$K_{\tau}$}

\begin{document}
\draft

\wideabs{
\title{
An Improved Limit on the Rate of the Decay
\Kpmel
 }
Submitted to Phys. Rev. Lett, 27 April, 2000\\

\author{
R. Appel$^{6,3}$, G.S. Atoyan$^4$, B. Bassalleck$^2$,
D.R. Bergman$^6$\cite{DB}, D.N. Brown$^3$\cite{DNB}, N. Cheung$^3$,
S. Dhawan$^6$,  \\
H. Do$^6$, J. Egger$^5$, S. Eilerts$^2$\cite{SE}, C. Felder$^{3,1}$,
H. Fischer$^2$\cite{HF}, M. Gach$^3$\cite{MG}, W. Herold$^5$, \\
V.V. Issakov$^4$, H. Kaspar$^5$, D.E. Kraus$^3$, D. M. Lazarus$^1$,
L. Leipuner$^1$, P. Lichard$^3$, J. Lowe$^2$, J. Lozano$^6$\cite{JL},\\
 H. Ma$^1$, W. Majid$^6$\cite{WMa}, W. Menzel$^7$\cite{WMe},
S. Pislak$^{8,6}$, A.A. Poblaguev$^4$,
V.E. Postoev$^4$, A.L. Proskurjakov$^4$, \\ P. Rehak$^1$,
P. Robmann$^8$, A. Sher$^3$, J.A. Thompson$^3$,
P. Tru\"ol$^{8,6}$, H. Weyer$^{7,5}$, and M.E. Zeller$^6$   \\
}

\address{
$^1$ Brookhaven National Laboratory, Upton L. I., NY 11973, USA\\
$^2$Department of Physics and Astronomy,
University of New Mexico, Albuquerque, NM 87131, USA\\
$^3$ Department of Physics and Astronomy, University of Pittsburgh,
Pittsburgh, PA 15260, USA \\
$^4$Institute for Nuclear Research of Russian Academy of Sciences,
Moscow 117 312, Russia \\
$^5$Paul Scherrer Institut, CH-5232 Villigen, Switzerland\\
$^6$ Physics Department, Yale University, New Haven, CT 06511, USA\\
$^7$Physikalisches Institut, Universit\"at Basel, CH-4046 Basel,Switzerland\\
$^8$ Physik-Institut, Universit\"at Z\"urich, CH-8057 Z\"urich, Switzerland}
\date{April 27, 2000}
\maketitle 
\begin{abstract}
We report results of a search for the lepton-family number violating 
decay \Kpmel\/ from
data collected by experiment E865 in 1996
at the Alternating Gradient Synchroton of Brookhaven National Laboratory.
We place an
upper limit on the branching ratio at $3.9 \times
10^{-11}$ (90\% C.L.). Together with results based on data
collected in 1995 and
an earlier experiment, E777, this result establishes a combined 90\% 
confidence level upper limit on the branching ratio at $2.8 \times 10^{-11}$.
We also report a new upper limit on the branching ratio for $\pi^0\rightarrow
\mu^+ e^-$ of $3.8\times 10^{-10}$ (90\% C.L.).
\end{abstract}

}

Despite the success of the Standard Model in describing elementary
particle
physics, several key issues remain unresolved.
One  is the gauge-hierarchy problem, {\it i.e.}, that the scalar Higgs
field,
introduced to give mass to the W and Z vector bosons,
suggests a particle of mass $\approx$ 100 GeV/$c^2$ while its
renormalization implies a mass scale at least ten orders of magnitude
larger.  Theoretical extensions to the Standard Model, such as
Horizontal Gauge models~\cite{Cahn80,Shanker81},
Technicolor~\cite{Susskind81}, and Supersymmetry~\cite{SUSY}, were
developed primarily to address the
gauge hierarchy problem.  These extensions   permit
lepton-family number (LFN) non-conserving decays.
Observation of such phenomena would indicate new physics: physics
beyond the Standard Model.

 To test these
theories in the kaon sector several experiments have recently been performed:
$K^{0}_{L} \rightarrow \mu^{\pm} e^{\mp}$\cite{E791_98},
$K^{0}_{L} \rightarrow \pi^{0} \mu^{\pm} e^{\mp}$\cite{E799_98}\ and
\Kpmel(\Kpme).
In the mid and late 1980s an experiment at the Brookhaven AGS, BNL E777,
searched
for the \Kpme\/ decay. Prior to that experiment the upper limit on the
branching ratio for \Kpme\/ was measured at CERN to be $4 \times
10^{-9}$~\cite{cern_pme}. E777 reduced that limit to $2.1 \times 10^{-10}$
(90\% C.L.)~\cite{campagnari}. In 1992  a more sensitive experiment,
BNL E865,  continued  the search for the \Kpme\/ mode. In the 1995 running 
period E865 achieved the same limit as E777 \cite{Bergman,Pislak}.
In this paper, we report new results,  from data collected in 1996.

\begin{figure}[htb]
\epsfig{figure=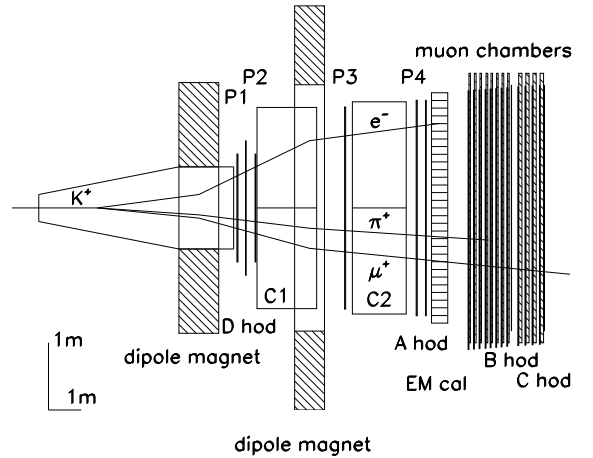,width=88mm}%
\centering
\caption{Plan view of the E865 apparatus.}%
\label{fig:detector}
\end{figure}

The E865  detector system and trajectories from a simulated \Kpme\ decay 
are shown in Fig.~\ref{fig:detector}. Some of the details of the
detector have been previously discussed in the context of an associated
measurement of the branching ratio \kpeel~\cite{piee}.

\vspace{-8mm}

The apparatus resided in an unseparated beam directly downstream from
a 5 m long vacuum decay volume.  With $10^{13}$ protons impinging on a 10 cm
long Cu production target, a secondary unseparated 6 GeV/c
beam was produced containing about $10^{8}\ K^+$ and
$2\times 10^{9} ~\pi^{+}$ and protons per 1.6 second AGS pulse.  All detector
elements were either desensitized or removed from the beam region.   A dipole 
magnet at the exit of the
decay region approximately separated $K^+$ decay products  by charge
(negative to beam left, positive to beam right) and reduced the charged
particle background originating upstream of the detector.  Proportional
chamber packages (P1-P4), each containing four planes of chambers,  were
arrayed on either side of a second dipole magnet to form a momentum analyzing
spectrometer system.  The momentum resolution of this configuration was
$\sigma_P\approx 0.003P^2$ GeV/$c$, where the momentum of the decay products, 
$P$,  ranged from 0.6 to 4 GeV/$c$.

Correct particle identification
(PID) was of critical importance in reducing backgrounds.  The first elements of
the PID system were atmospheric pressure \v{C}erenkov counters 
upstream and downstream of the spectrometer magnet.  
The left sides of each, C1L and C2L, were filled
with hydrogen gas to detect $e^-$ and not $\pi^-$, and had a light 
yield of $\approx$ 1.7 photoelectrons (p.e.) for $e^{-}$.
To reject $e^+$, the right sides (C1R and C2R) were filled with methane with 
a light yield of about 4.5 p.e. for $e^{+}$.  The two sides were separated 
by a thin membrane. In order to reduce beam gas interactions, closed tubes of 
hydrogen gas were placed in the beam region of C1R and C2R. 

The second PID element  was a Shashlyk-style
electromagnetic calorimeter (EM cal) ~\cite{Shashlyk}.  This device consisted of 600 
modules, each 11.4 by 11.4 cm$^2$ in cross section and 30 cm (15 radiation 
lengths) along
the beam, arrayed 30 horizontally by 20 vertically with 18 modules removed in 
the beam region.    The approximate
resolution of the array for electrons was $\sigma_E /E = 0.09/\sqrt{E}$, where
$E$ is the electron energy measured in GeV. Typical energy deposition of 
 a minimum ionizing particle was 250 MeV.  

The third element was a muon range stack with 24 planes of proportional tubes, 
with tubes alternately oriented horizontally and vertically, and plates of 
steel placed between each pair of planes.  The steel thickness was 5 cm 
between the first eight pairs of planes and 10 cm between the last four.

Trigger hodoscopes were located directly downstream of P1 (D hod),
immediately upstream of the calorimeter (A hod), and between the eighth and
ninth pairs of proportional tubes in the range stack (B hod).

The first-level trigger (T0) selected three charged particle tracks 
based on hodoscope and calorimeter hit patterns consistent with kaon three 
body decays, and with at least one particle on each side of the apparatus.
The $\pi \mu e$ trigger added a signal from  B hod, and at least 0.25 p.e. 
from C1L and C2L.  The T0 trigger, downscaled by $ 10^{4}$, also served as the 
trigger for our normalizing process,  
$K^+\rightarrow \pi^+ \pi^+ \pi^-\ (K_{\tau})$.
Satisfaction of any requested
trigger, roughly 700 times per machine pulse, resulted in all information in 
the various data buffers being read out to a Fastbus based data acquisition 
system in about 100 $\mu$s.

This paragraph describes the most important offline selection 
criteria \cite{Do}.
Common requirements for the \Kpme\/ and the K$_{\tau}$ normalizing mode included
a vertex formed from three reconstructed tracks, one negatively 
and two positively charged, with total vector momentum consistent with the 
beam phase space.  Events with an extra (accidental) track on the left, not 
on the vertex, capable of making a C1L or C2L signal were removed.
For  \Kpme\/ candidates, additional cuts were required. 
Background from $\pi^{0} \rightarrow \gamma e^{+}e^{-}$ (Dalitz) was 
suppressed  by removing
events with $M_{ee}$ (effective mass of the $e^{-}$ with either positive 
particle interpreted as $e^{+}$ ) $<$ 50 Mev/c$^{2}$.
Electron PID required a signal in both C1L and C2L with 
corrected timing within $\pm$ 4 ns, and energy deposited in the calorimeter
divided by the particle measured momentum ($E/P$) to be 1.0 $\pm$ 0.2.
Pions were required to have C1R and C2R $<$ 1.2 p.e., $E/P$ $<$ 0.85, and to 
not have penetrated the range stack to a depth of their 
expected ionization loss range. 
Muons were required to have C1R and C2R $<$ 1.2 p.e., less than 450 MeV 
deposited in the calorimeter, and range stack penetration depth consistent
with their range.  High rates on the right, mostly at the one p.e. level, from
high momentum muons and charged particle scintillation, prevented use of lower
 $\pi$ and $\mu$ thresholds in C1R and C2R.
Table I summarizes the final PID efficiencies and probabilities of
misidentification \cite{Do}.  These measurements were made with particles of 
known identities from \kppp\/ and from $K_{\pi 2}$ and $K_{\mu 3}$ with a 
subsequent Dalitz decay of the $\pi^0$.

\vspace{-3mm}
\begin{flushleft}
\begin{tabular}{c|c|c|c} \hline\hline
\hspace{.43in} & $\rightarrow \pi$ &
                $\rightarrow \mu$ &
                $\rightarrow e$                \\ \hline
                $\pi^+$                          &
                $0.780 \pm 0.004$                  &
                $0.049 \pm 0.017$                  &
                 -                                              \\ \hline
                $\pi^-$                           &
                $0.969 \pm 0.002$                  &
                 -                                 &
                $(2.6 \pm .1) \times 10^{-6}$                  \\ \hline
                $\mu^+$                           &
                 -                  &
                $0.743 \pm 0.014$                  &
                 -                                              \\ \hline
                $e^+$                             &
                $(1.7 \pm .7) \times 10^{-5}$                  &
                $<1.7 \times 10^{-5}$     &
                $0.873 \pm 0.002$                               \\ \hline
                $e^-$                             &
                 -                                 &
                 -                                 &
                $0.546 \pm 0.003$               \\ \hline\hline
\end{tabular}

\vspace{3mm}

TABLE I. PID efficiencies and probabilities of misidentification.
The symbol ``$\rightarrow$'' denotes ``identified as''.\\
\end{flushleft}

\vspace{-3mm}

A likelihood analysis was used to evaluate the probability 
that selected events fit particular hypotheses, {\it e.g.},  \Kpme\/,
$K_{\tau}$, accidental. In this method distributions derived from data  
were used as probability density functions (PDFs).  These distributions included
vertex and track quality, reconstructed beam parameters, PID detector responses 
and timing, and the invariant mass of the three decay products, to mention
the most important.  The extremes of these distributions 
were cut to allow the survival of about 95\% of the events for the respective 
hypotheses.  The resulting PDFs were then the templates to determine the 
probabilities that the variables in a given event originated
from the hypothesized mode, say the $i^{th}$ mode.  
The logarithms of these probabilities were added to form the 
joint log-likelihood ($\L_i$) for the $i^{th}$ decay hypothesis.  
In the case of \Kpme\/ the 
PDF for the invariant mass of the decay products was generated by Monte Carlo 
simulation.  All other PDFs were generated from data.

\begin{figure}[t]
\epsfig{figure=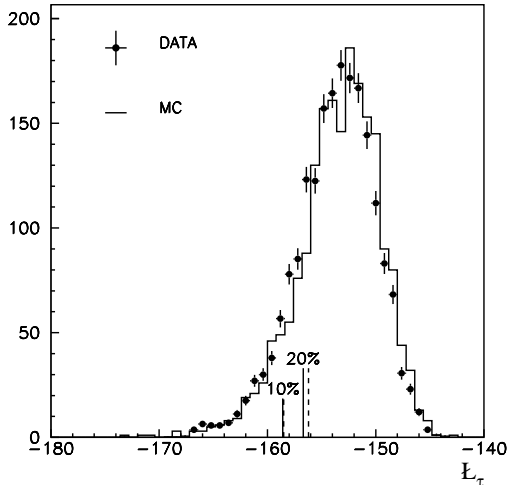,width=72mm}\centering
\caption{\kppp\/ log-likelihood comparison between data and
Monte-Carlo events. The vertical lines (solid for data and dash for MC) 
show the 10\% and 20\% likelihood points.}
\label{fig:tau_comp}
\end{figure}

An example of the final {$\L_\tau$} distribution is seen in 
Fig.~\ref{fig:tau_comp} where we display data and Monte Carlo 
simulated $K_{\tau}$ events.  The 10\% and 20\%  points on these plots 
represent 
likelihood values for which the probabilities of finding smaller likelihood 
are 10\% and 20\%, respectively. 
  
\begin{figure}[H]
\epsfig{figure=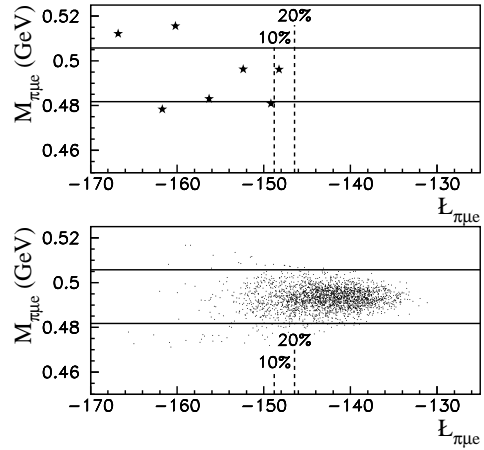,width=72mm}\centering
\caption{Scatter plot of E865 1996 \Kpme\/ data (top) and Monte Carlo
(bottom).  The abscissa is the
log-likelihood of the reconstructed events under the \Kpme\/ hypothesis, the
ordinate is
the invariant mass of the detected particles.  The horizontal lines demark the
3$\sigma$ mass region.}
\label{fig:pmelik_comp}
\end{figure}

Candidate \Kpme\/ events were first subjected to the cuts described above, and
for those that survived the $\L_{\pi \mu e}$ was determined. 
The results of this analysis are presented in Fig.~\ref{fig:pmelik_comp}
as a scatter plot of $\L_{\pi \mu e}$ {\em vs.} invariant mass of the decay 
products.  The top plot is data where the invariant mass cut was increased 
from its nominal value of 3$\sigma$ to 6$\sigma$ for display purposes.  The 
bottom plot shows simulated  \Kpme\/ events, where a vector interaction for the 
decay was assumed in the simulation.
Only seven data events have survived the cuts with a
$\L_{\pi \mu e}$ greater than -170.  Three of these are within the 
3$\sigma$ accepted \mpme\/ region, {\it i.e.}, between the horizontal lines. 
The three data events which pass all cuts have probabilities of 13\% (79\%), 
2\% (52\%), and 0.5\% (64\%), respectively, of being consistent with a 
\Kpme\/ (accidental) hypothesis. 
 
The three most probable sources of background events are 
\kppp\/, Dalitz, and accidentals. 
Contributions to
backgrounds from the \kppp\/ and Dalitz modes were from misidentification
of particle species and incorrect track reconstruction, while accidental 
backgrounds were from events with correct PID, but for which the
three particles did not originate from a single decay and occurred 
accidentally in time.  

Estimation of the number and $\L_{\pi \mu e}$ distribution for   
\kppp\/ and Dalitz events was accomplished using events with correct PID, and 
replacing their detector response values with those 
corresponding to \Kpme\/. This replacement was made by selecting randomly from 
a library of measured responses for events where the respective detectors 
gave incorrect PID, 
{\it e.g.}, the  C1L response of a $\pi^-$ from a $K_{\tau}$ event was replaced
with one for which a known $\pi^-$ had an $e^-$ response in C1L.
The probability of such misidentification could 
then be calculated, and the number
of such events normalized to the measured number of \kppp\/ and Dalitz events
in the total data sample.  

\vspace{-6mm}

The number and $\L_{\pi \mu e}$ distribution of accidental events was estimated
using events for which the variable describing the rms deviation
from the mean time of all participating counters ($T_{rms}$) was more than 
3 standard deviations, but which otherwise satisfied all \Kpme\/ cuts.
Estimation of the relative probability that such 
events would have an acceptable $T_{rms}$ was made by evaluating the
$T_{rms}$ distribution for events with total momentum greater than
6.5 GeV/c, {\it i.e.}, events which are primarily accidental.  The $T_{rms}$ 
distribution for high momentum events also gave the $T_{rms}$ 
PDF for accidental events with acceptable values of $T_{rms}$.  The 
latter was used to randomly replace the $T_{rms}$ value for accidental events
described above in forming the full $\L_{\pi \mu e}$ distribution for 
accidentals.  

\begin{figure}[t]
\epsfig{figure=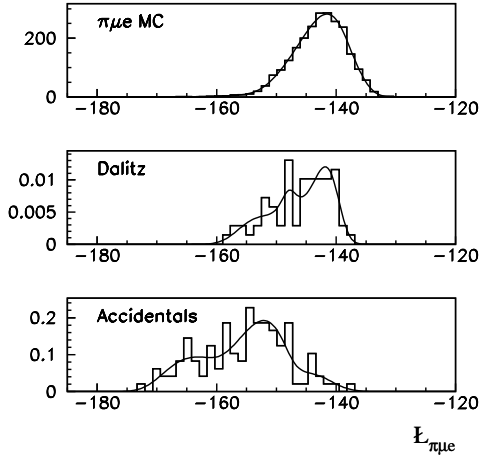,width=72mm}\centering
\caption[\Kpme\/ likelihood distributions for signal, Dalitz, and
accidentals] 
{\Kpme\/ likelihood distributions for signal, Dalitz, and
accidentals. The histograms are smoothed to compensate for the lack of 
statistics.  The vertical scales for Dalitz and accidental distributions are
normalized to their respective number of expected events.} 
\label{fig:pdf_modes}
\end{figure}

The $\L_{\pi \mu e}$ distributions for \Kpme\/, 
Dalitz, and accidental  modes are shown 
in Fig.~\ref{fig:pdf_modes}~\cite{tau_ll}.  The estimated number of
such events that would pass all \Kpme\/ cuts and appear in 
Fig.~\ref{fig:pmelik_comp} within the accepted $M_{\pi\mu e}$ region is 
$2.6\pm 1.0$: $0.06 \pm 0.03$ $K_{\tau}$ events,
$0.1 \pm 0.1$ Dalitz events, and $2.4 \pm 1.0$ accidental events, in good
agreement with the three events observed.

With the $\L_{\pi \mu e}$ distributions for \Kpme\/, and those of the most 
dominant background modes weighted in relative proportion as determined above,
a $\chi ^2$ function for Poisson-distributed data was 
minimized to determine the most probable number of \Kpme\/ and total background
events in the data distribution of Fig.~\ref{fig:pmelik_comp}.  Those numbers
were 0.0 and 3.0, respectively, consistent with our estimation of 2.6 background
events.  To determine the 90\% confidence interval for
this result, the Frequentist approach was used~\cite{Feldman98}, 
with the number of background 
events assigned to be 2.6.  Including the uncertainty in the assigned background
level, the result of that analysis was that
the expected number of \Kpme\/ events in our data sample 
is less than 2.5 at the 90\% confidence level.  

\vspace{-10mm}
\vspace{-2mm}

The upper limit on the \Kpme\/ branching ratio, normalized to the $K_{\tau}$
branching ratio, is calculated according to the formula:
\vspace{-1mm}
\[BR(\pi \mu e)<BR(\pi\pi\pi) \cdot
\frac{{\em N(\pi \mu e)}}{{\em N(\pi\pi\pi)}}\cdot
\frac{{\em A_{\pi\pi\pi}}}{{\em A_{\pi \mu e}}} \cdot
C,\]
where {\it BR} denotes the noted branching ratio, $BR(\pi\pi\pi)=0.0559 \pm
0.0005$~\cite{PDT}; $N(\pi \mu e)=2.5$, the 90\% C.L.
number of
signal events; $N(\pi \pi \pi)=2.19 \times 10^{10}$, the number of
\kppp\/ events adjusted
for downscale factors;
$A$  represents the geometrical acceptance of the detector system for
the specific decay mode, $A_{\pi\pi\pi}/A_{\pi \mu e} = 1.64 \pm 0.02$, 
with 0.01 contribution from systematic uncertainty; 
and $C  = 3.78 \pm 0.08$, the product of correction factors accounting for
efficiency differences between the two modes.
The bulk of $C$, 3.15, is the reciprocal of the product 
of the $\pi^+,\mu^+,e^-$ PID efficiencies shown in 
Table I, while the remaining 1.20 results from acceptance differences between
\Kpme\/ and $K_{\tau}$ due to cuts~\cite{Do}.

Employing these factors, we set a limit on the branching
ratio $BR(\pi \mu e) < 3.9 \times 10^{-11}$ (90\% C.L.). Combining
this result with data collected in 1995,
$BR< 2.1 \times 10^{-10}$~\cite{Bergman,Pislak}, and
the E777 experiment, $BR< 2.0 \times 10^{-10}$~\cite{campagnari},
yields a new upper limit of the branching ratio for \Kpme\/ of 
$2.8 \times 10^{-11}$ (90\% C.L.).

This branching ratio implies that an intermediate boson in models described 
by a horizontal
gauge interaction, {\it e.g.}, Ref.~\cite{Cahn80,Shanker81},  with
purely vector coupling and strength equal to that of the weak interaction,
would have a mass greater than 60 TeV.

Since the process $\pi^0\rightarrow \mu^+ e^-$  
would be observed in our data through $K^+\rightarrow \pi^+ \pi^0;
\ \pi^0\rightarrow \mu^+ e^-$, we also set an upper limit on
its branching ratio.  The only candidate events  are the three discussed 
above, but their $M_{\mu e}$ values of 0.226, 0.282, and 0.332 GeV/$c^2$ are 
too far from $M_{\pi^0}$ for the events to have originated from $\pi^0$ decays.
We thus place an upper limit on the expected number of 
$\pi^{0} \rightarrow \mu^{+} e^{-}$ events at 2.44 (90\% C.L.).  
We again normalize to the $K_{\tau}$ mode with the ratio of acceptances being
$A_{\pi\pi\pi}/A_{\pi2(\mu e)}=4.07\pm 0.02$, the factor C=$3.22\pm 0.07$,  and 
the branching ratio for $K_{\pi2}$ is 21.16\%~\cite{PDT}.  Combining these 
factors gives an upper limit on the decay branching ratio for 
$\pi^0\rightarrow \mu^+ e^-$ of $3.8\times 10^{-10}$ (90\% C.L.), 
compared with the Review of Particle Physics limit of $1.72\times 10^{-8}$ 
for the combined $\pi^0\rightarrow (\mu^+e^- + \mu^-e^+)$ decays~\cite{PDT}. 

We gratefully acknowledge  contributions to the success of
this experiment by 
the staff and management of the AGS at the Brookhaven National
Laboratory, and the technical staffs of the participating institutions.
This work was supported in part by the U. S. Department of Energy,
the National Science Foundations of the USA(REU program), Russia and Switzerland, and
the Research Corporation.


\begin{references}
\bibitem{tau_ll}There were insufficient $K_{\tau}$ events with 
acceptable \Kpme\/ kinematics to derive a $\L$ distribution for 
that mode.
\end{references}
\end{document}